=======  **INFORMATION TECHNOLOGY IN ENGINEERING SYSTEMS**  =======

# Composition of Management System for Smart Homes
# ("Information Processes", 2010, 10(1), 78-86)

## Mark Sh. Levin*, Aliaksei Andrushevich**, Alexander Klapproth**

*\*Inst. for Information Transmission Problems, Russian Academy of Sciences,*
*19 Bolshoj Karetny lane, Moscow 127994, Russia*
*email: mslevin@acm.org*
*\*\*CEESAR-iHomeLab, Lucerne University of Applied Sciences,*
*Technikumstrasse 21, CH-6048 Horw, Switzerland*
*email: {aliaksei.andrushevich,alexander.klapproth}@hslu.ch*

**Abstract**—The paper addresses modular hierarchical design (composition) of a management system for smart homes. The management system consists of security subsystem (access control, alarm control), comfort subsystem (temperature, etc.), intelligence subsystem (multimedia, houseware). The design solving process is based on Hierarchical Morphological Multicriteria Design (HMMD) approach: (1) design of a tree-like system model, (2) generation of design alternatives for leaf nodes of the system model, (3) Bottom-Up process: (i) multicriteria selection of design alternatives for system parts/components and (ii) composing the selected alternatives into a resultant combination (while taking into account ordinal quality of the alternatives above and their compatibility). A realistic numerical example illustrates the design process of a management system for smart homes.

## 1. INTRODUCTION

In recent years smart homes (i.e., home environment management systems) are increasing in popularity. For example, the following research directions have been intensively studied: (1) general issues and architecture for smart homes ([10], [11], [12], [15], [19], [27], [30], [31], [44], [48] [47]); (2) design issues ([3], [10], [15], [37]); (3) management issues ([2], [5], [11], [13], [27], [33], [45]); (4) security issues ([17], [18]); (5) dynamics issues of smart home [32]; (6) wireless issues ([26], [40]); (7) sensor systems ([8], [29]); and (8) data integration/fusion ([14], [45]). In the article, modular hierarchical design (composition) of a management system for smart homes is firstly suggested. The design approach consists in a modular composition of a configuration for an electronic system consisting of several parts/subsystems (e.g., security subsystem, comfort subsystem, intelligence subsystem) (Fig. 1).

Various approaches have been applied for the design of system configurations: (1) the shortest path problem [1]; (2) evolutionary approaches (genetic algorithms, etc.) (e.g., [34]); (3) multi-agent approaches (e.g., [7]); (4) approaches based on fuzzy sets (e.g., [39]); (5) constraint-based methods (e.g., [4], [34]) including composite constraint satisfaction problems (e.g., [36], [42]); (6) ontology-based approaches (e.g., [9]); (7) multiple choice knapsack problem (e.g., [16]); (8) multicriteria multiple choice knapsack problem (e.g., [25], [41]); (9) hierarchical multicriteria morphological design (HMMD) approach ([20], [21], [22]); (10) AI techniques (e.g., [28], [43], [46]); and (11) design grammars approaches (e.g., multidisciplinary grammar approach that includes production rules and optimization, graph grammar approach) (e.g., [38]). A survey of combinatorial optimization approaches to system configuration design is presented in [23].

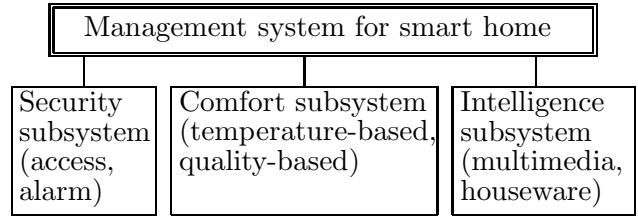

Fig. 1. Structure of management system

In this article HMMD approach is used which implements a hierarchical modular system design. The approach is based on two optimization problems: (i) multicriteria ranking (outranking technique as modification of ELECTRE [35]) and (ii) morphological synthesis based on morphological clique problem ([20], [21], [22]). Development history of morphological approaches is presented in [24]. HMMD implements a multi-stage design framework and provides the following ([20], [21], [22]): (1) cascade-like design framework: (i) decomposition/partitioning of system and system requirements to obtain a hierarchical system model and a hierarchy of system requirements which correspond to system parts/components, (ii) Bottom-Up design process; (2) parallel and independent assessment and analysis of design alternatives for system parts/ components; (3) integration of analytical, computer-based, and expert-based evaluation support procedures; (4) parallel analysis and design (evaluation, selection, composition) of design alternatives for composite system parts/components; and (5) opportunity to use cognitive methods at each step and/or part of the design process.

Our realistic numerical design example involves hierarchical (tree-like) structure of a management system for smart homes, design alternatives (DAs) for the system parts/components, and Bottom-Up solving process. Expert judgement is used for assessment of DAs and their compatibility. All estimates and computing process are only illustrative but may be used as a basis for real world applications.

## 2. UNDERLAYING PROBLEMS

First, let a consider multicriteria ranking. Let $H = \{1, ..., i, ..., t\}$ be a set of items which are evaluated upon criteria $K = \{1, ..., j, ..., d\}$ and $z_{i,j}$ is an estimate (quantitative, ordinal) of item $i$ on criterion $j$. The matrix $\{z_{i,j}\}$ is a basis to build a partial order on $H$, for example through the following generalized scheme: (a) pairwise elements comparison to get a preference (and/or incomparability, equivalence) binary relation, (b) building a partial order on $H$. Here the following partial order (partition) as linear ordered subsets of $H$ is searched for: $H = \cup_{k=1}^{m} H(k)$, $|H(k_1) \cap H(k_2)| = 0$ if $k_1 \neq k_2$, $i_2 \preceq i_1$ $\forall i_1 \in H(k_1)$, $\forall i_2 \in H(k_2)$, $k_1 \leq k_2$. Set $H(k)$ is called layer $k$, and each item $i \in H$ gets priority $r_i$ that equals the number of the corresponding layer. The list of basic techniques for multicriteria selection/ranking is the following [6]: (1) multi-attribute utility analysis; (2) multi-criterion decision making; (3) Analytic Hierarchy Process (AHP); and (4) outranking techniques. In the article, a version of outranking technique is used [35].

Hierarchical Morphological Multicriteria Design (HMMD) based of morphological clique problem was firstly suggested in 1994 and is described in ([20], [21], [22]). Here a composite (modular, decomposable) system under examination consists of the components and their interconnections or compatibilities. Basic assumptions of HMMD are the following: (a) a tree-like structure of the system; (b) a composite estimate for system quality that integrates components (subsystems, parts) qualities and qualities of interconnections (hereinafter referred as 'IC') across subsystems; (c) monotonic criteria for the system and its components; and (d) quality of system components and IC are evaluated on the basis of coordinated ordinal scales. The designations are: (1) design alternatives (DAs) for leaf nodes of the model; (2) priorities of DAs ($r = \overline{1, k}$; 1 corresponds to the best one); (3) ordinal compatibility (IC) for each pair of DAs ($w = \overline{0, l}$; $l$ corresponds to the best one). The basic phases of HMMD are:

*Phase 1.* Design of the tree-like system model (a preliminary phase).

*Phase 2.* Generating DAs for model's leaf nodes.

*Phase 3.* Hierarchical selection and composing of DAs into composite DAs for the corresponding higher level of the system hierarchy (morphological clique problem).

*Phase 4.* Analysis and improvement of the resultant composite DAs (decisions).

Let $S$ be a system consisting of $m$ parts (components): $P(1), ..., P(i), ..., P(m)$. A set of design alternatives is generated for each system part above. The problem is:

*Find a composite design alternative $S = S(1) \star ... \star S(i) \star ... \star S(m)$ of DAs (one representative design alternative $S(i)$ for each system component/part $P(i)$, $i = \overline{1, m}$) with non-zero IC between design alternatives.*

A discrete space of the system excellence on the basis of the following vector is used: $N(S) = (w(S); n(S))$, where $w(S)$ is the minimum of pairwise compatibility between DAs which correspond to different system components (i.e., $\forall P_{j_1}$ and $P_{j_2}$, $1 \le j_1 \ne j_2 \le m$) in $S$, $n(S) = (n_1, ..., n_r, ...n_k)$, where $n_r$ is the number of DAs of the $r$th quality in $S$. As a result, we search for composite decisions which are nondominated by $N(S)$ (Fig. 2 and Fig. 3).

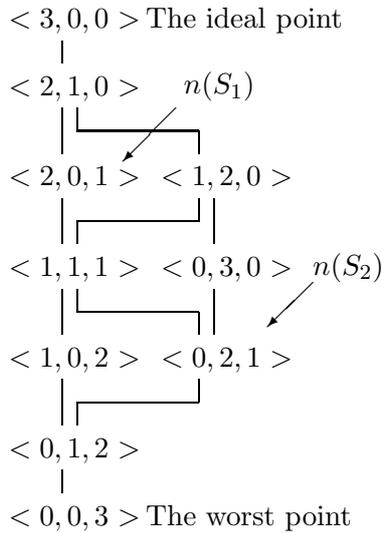

Fig. 2. Lattice of system quality

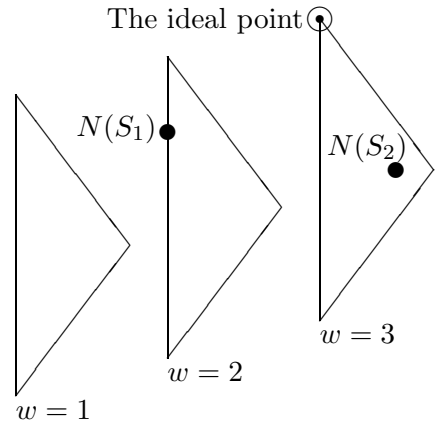

Fig. 3. Space of system quality by $N(S)$

Fig. 4 and Fig. 5 illustrate the composition problem. Here examples of composite solutions are (Fig. 2, Fig. 3, Fig. 4, and Fig. 5): $S_1 = X_2 \star Y_1 \star Z_2$, $N(S_1) = (2; 2, 0, 1)$; $S_2 = X_1 \star Y_2 \star Z_3$, $N(S_2) = (3; 0, 2, 0)$.

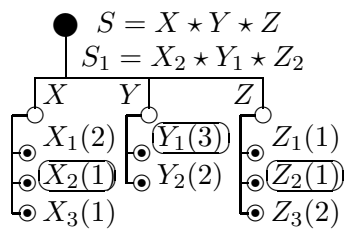

Fig. 4. Example of composition

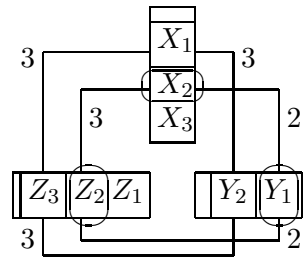

Fig. 5. Concentric presentation

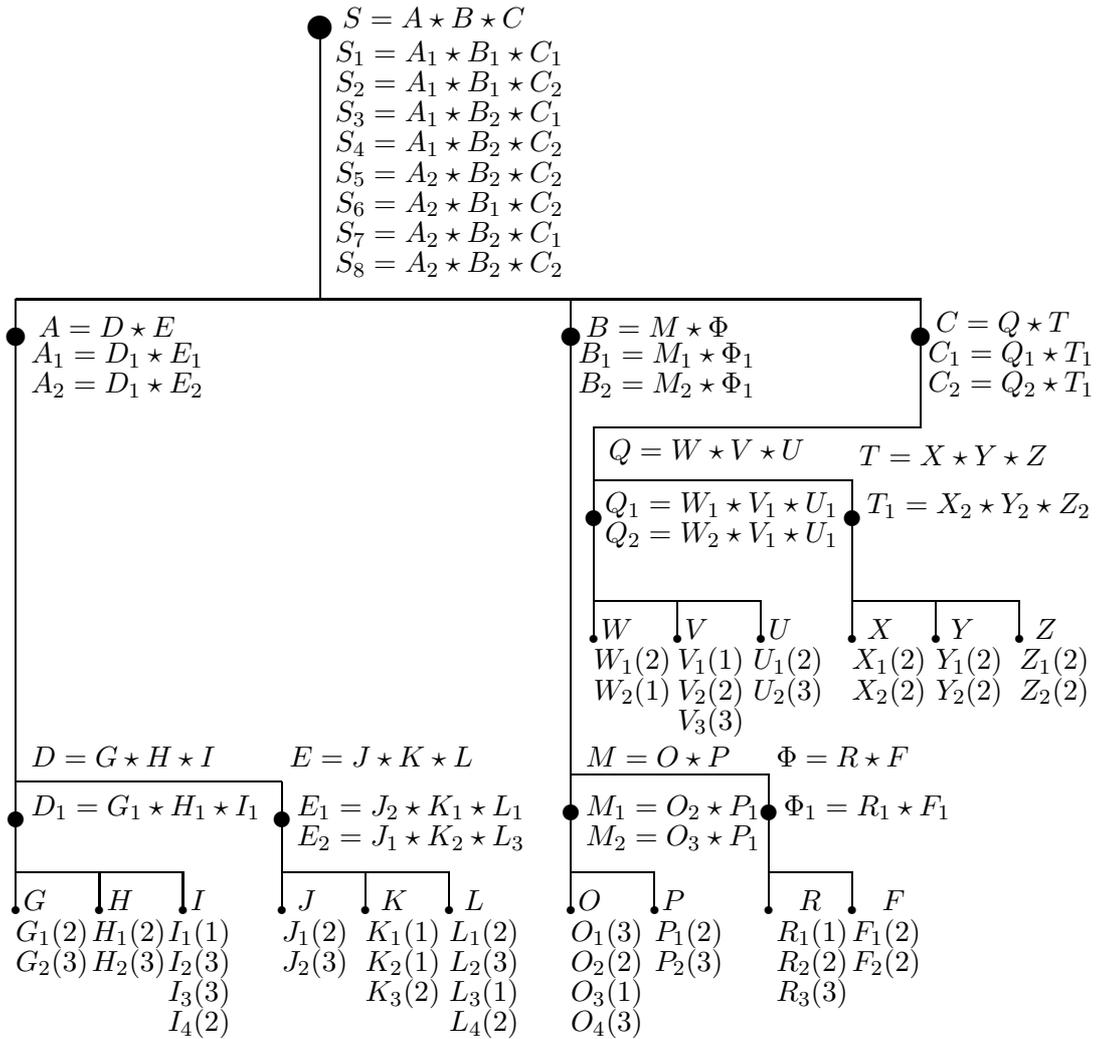

Fig. 6. Hierarchical structure of management system for smart homes

## 3. HIERARCHICAL MODEL AND COMPONENTS

The following hierarchical structure of management system for smart home is examined (Fig. 6):

**0.** Management system $S = A \star B \star C$.

**1.** Security subsystem $A = D \star E$.

    *1.1.* Access control $D = G \star H \star I$.

        *1.1.1.* Windows shutters $G$: Manual $G_1$, Electricity-driven $G_2$.

        *1.1.2.* Door locks $H$: Standard $H_1$, Electric $H_2$.

        *1.1.3.* Authentication point $I$: Physical key $I_1$, PIN $I_2$, RFID $I_3$, Biometric $I_4$.

    *1.2.* Alarm control $E = J \star K \star L$.

        *1.2.1.* Alarm signal $J$: Buzzer $J_1$, Light $J_2$.

        *1.2.2.* Presence detector $K$: Infrared $K_1$, Ultrasonic $K_2$, Motion $K_3$.

        *1.2.3.* Alert connection $L$: Landline $L_1$, Radio $L_2$, Internet $L_3$, GSM/SMS $L_4$.

**2.** Comfort subsystem $B = M \star N$.

*2.1.* Temperature-based    $M = O \star P$.

  *2.1.1.* Heating    $O$: Floor $O_1$, Radiators $O_2$. Roof $O_3$, Thermo-wall $O_4$.

  *2.1.2.* Air-conditioning    $P$: External $P_1$, Internal $P_2$.

*2.2.* Quality-based    $N = R \star F$.

  *2.2.1.* Ventilation fan    $R$: Ceiling $R_1$, Working places $R_2$. Central $R_3$.

  *2.2.2.* Air filter    $F$: Oven-based $F_1$, Central-based $F_2$.

**3.** Intelligence subsystem    $C = Q \star T$.

*3.1.* Multimedia    $Q = W \star V \star U$.

  *3.1.1.* Video-system    $W$: Monitor $W_1$, Beamer $W_2$.

  *3.1.2.* Audio-system    $V$: "2:1" $V_1$, "5:1" $V_2$, Dolby $V_3$.

  *3.1.3.* Home server / PC    $U$: Decoupled $U_1$, Integrated $U_2$.

*3.2.* Houseware    $T = X \star Y \star Z$.

  *3.2.1.* Oven    $X$: Gas $X_1$, Electric $X_2$.

  *3.2.2.* Refrigerator    $Y$: With freezer $Y_1$, Web-enabled $Y_2$.

  *3.2.3.* Vacuum cleaner    $Z$: Central $Z_1$, iLoc-enabled $Z_2$.

### *3.1. Assessment and Priorities*

The following criteria are used for assessment of DAs ('+' corresponds to positive orientation of an ordinal scale as [1,5] and '-' corresponds to the negative orientation of the scale): *(a)* cost $C_1, '-'$; *(b)* energy consumption $C_2, '-'$; *(c)* reliability $C_3, '+'$; and *(d)* life cycle length $C_4, '+'$. Criteria weights for six system parts are contained in Table 1.

Table 2. Estimates

| DAs | Criteria | | | | Priority |
|-----|---|---|---|---|---|
| | 1 | 2 | 3 | 4 | |
| $G_1$ | 1 | 0 | 3 | 3 | 2 |
| $G_2$ | 3 | 2 | 3 | 2 | 3 |
| $H_1$ | 1 | 0 | 3 | 3 | 2 |
| $H_2$ | 3 | 2 | 3 | 2 | 3 |
| $I_1$ | 1 | 0 | 3 | 4 | 1 |
| $I_2$ | 2 | 1 | 3 | 3 | 3 |
| $I_3$ | 3 | 2 | 4 | 4 | 3 |
| $I_4$ | 4 | 3 | 5 | 5 | 2 |
| $J_1$ | 2 | 2 | 2 | 3 | 2 |
| $J_2$ | 3 | 1 | 2 | 3 | 3 |
| $K_1$ | 2 | 2 | 3 | 3 | 1 |
| $K_2$ | 2 | 2 | 3 | 3 | 1 |
| $K_3$ | 3 | 3 | 3 | 3 | 2 |
| $L_1$ | 1 | 1 | 2 | 2 | 2 |
| $L_2$ | 2 | 2 | 3 | 3 | 3 |
| $L_3$ | 2 | 2 | 4 | 3 | 1 |
| $L_4$ | 3 | 3 | 4 | 4 | 2 |
| $O_1$ | 3 | 2 | 2 | 2 | 3 |
| $O_2$ | 1 | 3 | 4 | 4 | 2 |
| $O_3$ | 2 | 2 | 3 | 2 | 1 |
| $O_4$ | 3 | 3 | 2 | 2 | 3 |

Table 3. Estimates

| DAs | Criteria | | | | Priority |
|-----|---|---|---|---|---|
| | 1 | 2 | 3 | 4 | |
| $P_1$ | 2 | 2 | 3 | 3 | 2 |
| $P_2$ | 4 | 4 | 3 | 2 | 3 |
| $R_1$ | 2 | 2 | 3 | 3 | 1 |
| $R_2$ | 3 | 3 | 2 | 3 | 2 |
| $R_3$ | 4 | 4 | 1 | 1 | 3 |
| $F_1$ | 2 | 2 | 1 | 2 | 2 |
| $F_2$ | 4 | 4 | 3 | 3 | 2 |
| $W_1$ | 4 | 3 | 3 | 3 | 2 |
| $W_2$ | 2 | 1 | 3 | 3 | 1 |
| $V_1$ | 1 | 1 | 3 | 3 | 1 |
| $V_2$ | 2 | 2 | 3 | 3 | 2 |
| $V_3$ | 3 | 3 | 3 | 3 | 3 |
| $U_1$ | 3 | 3 | 4 | 3 | 2 |
| $U_2$ | 4 | 3 | 3 | 3 | 3 |
| $X_1$ | 2 | 1 | 2 | 3 | 2 |
| $X_2$ | 3 | 3 | 3 | 3 | 2 |
| $Y_1$ | 2 | 3 | 3 | 3 | 2 |
| $Y_2$ | 3 | 2 | 3 | 3 | 2 |
| $Z_1$ | 3 | 3 | 3 | 3 | 2 |
| $Z_2$ | 2 | 2 | 2 | 3 | 2 |

Table 1. Criteria weights

| System component | Criteria | | | |
|-----|---|---|---|---|
| | 1 | 2 | 3 | 4 |
| 1. $D$ | $-2$ | $-1$ | 2 | 3 |
| 2. $E$ | $-1$ | $-1$ | 3 | 3 |
| 3. $M$ | $-2$ | $-2$ | 3 | 3 |
| 4. $\Phi$ | $-2$ | $-2$ | 2 | 2 |
| 5. $Q$ | $-3$ | $-3$ | 1 | 1 |
| 6. $T$ | $-1$ | $-1$ | 3 | 3 |

Tables 2 and 3 contain ordinal estimates of DAs upon the above-mentioned criteria (expert judgment) and priorities of DAs (as a result of multicriteria ranking based on Electre-like method). Priorities of DAs are shown in Fig. 6 (in parentheses) as well.

Estimates of compatibility between DAs are contained in Tables 4, 5, 6, 7, 8, and 9 (expert judgment).

Table 4. Compatibility

|  | $H_1$ | $H_2$ | $I_1$ | $I_2$ | $I_3$ | $I_4$ |
|---|---|---|---|---|---|---|
| $G_1$ | 3 | 3 | 3 | 2 | 1 | 1 |
| $G_2$ | 3 | 3 | 3 | 3 | 3 | 3 |
| $H_1$ |  |  | 3 | 1 | 1 | 1 |
| $H_2$ |  |  | 1 | 3 | 3 | 3 |

Table 5. Compatibility

|  | $K_1$ | $K_2$ | $K_3$ | $L_1$ | $L_2$ | $L_3$ | $L_4$ |
|---|---|---|---|---|---|---|---|
| $J_1$ | 2 | 1 | 3 | 2 | 1 | 1 | 3 |
| $J_2$ | 3 | 3 | 3 | 3 | 3 | 3 | 2 |
| $K_1$ |  |  |  | 3 | 2 | 0 | 2 |
| $K_2$ |  |  |  | 2 | 1 | 1 | 2 |
| $K_3$ |  |  |  | 2 | 3 | 2 | 2 |

Table 6. Compatibility

|  | $O_1$ | $O_2$ | $O_3$ | $O_4$ |
|---|---|---|---|---|
| $P_1$ | 3 | 3 | 2 | 1 |
| $P_2$ | 2 | 3 | 1 | 2 |

Table 8. Compatibility

|  | $V_1$ | $V_2$ | $V_3$ | $U_1$ | $U_2$ |
|---|---|---|---|---|---|
| $W_1$ | 3 | 2 | 1 | 3 | 2 |
| $W_2$ | 1 | 2 | 3 | 2 | 3 |
| $V_1$ |  |  |  | 3 | 1 |
| $V_2$ |  |  |  | 3 | 2 |
| $V_3$ |  |  |  | 2 | 3 |

Table 9. Compatibility

|  | $Y_1$ | $Y_2$ | $Z_1$ | $Z_2$ |
|---|---|---|---|---|
| $X_1$ | 2 | 2 | 3 | 2 |
| $X_2$ | 3 | 3 | 2 | 3 |
| $Y_1$ |  |  | 3 | 2 |
| $Y_2$ |  |  | 3 | 3 |

Table 7. Compatibility

|  | $R_1$ | $R_2$ | $R_3$ |
|---|---|---|---|
| $F_1$ | 3 | 3 | 2 |
| $F_2$ | 2 | 2 | 3 |

### 3.2. Composite Decisions

Now let us consider composite DAs. The following Pareto-efficient composite DAs are obtained for components of part $A$: $D_1 = G_1 \star H_1 \star I_1$, $N(D_1) = (3;1,2,0)$; $E_1 = J_2 \star K_1 \star L_1$, $N(E_1) = (3;1,1,1)$; $E_2 = J_1 \star K_2 \star L_3$, $N(E_2) = (1;1,2,0)$. Fig. 7 illustrates the space of system quality for $E$.

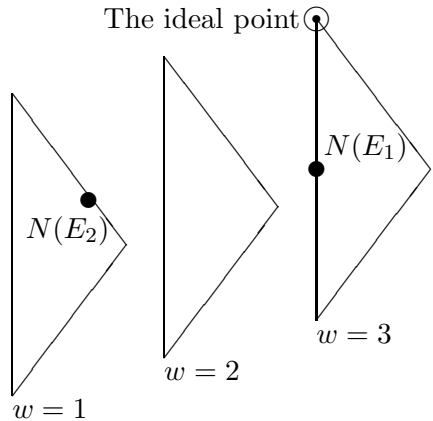

Fig. 7. Space of system quality for $E$

Thus the following composite DAs are obtained for part $A$: $A_1 = D_1 \star E_1$, $A_2 = D_1 \star E_2$.

The following Pareto-efficient composite DAs are obtained for components of part $B$:

$\Phi_1 = R_1 \star F_1$, $N(\Phi_1) = (3;1,1,0)$; $M_1 = O_2 \star P_1$, $N(M_1) = (3;0,2,0)$;

$M_2 = O_3 \star P_1$, $N(M_2) = (2;1,1,0)$.

Thus the following composite DAs are obtained for part $B$: $B_1 = \Phi_1 \star M_1$, $B_2 = \Phi_1 \star M_2$.

The following Pareto-efficient composite DAs are obtained for components of part $C$:

$Q_1 = W_1 \star V_1 \star U_1$, $N(Q_1) = (3; 1, 2, 0)$;    $Q_2 = W_2 \star V_1 \star U_1$, $N(Q_2) = (1; 2, 1, 0)$;

$T_1 = X_2 \star Y_2 \star Z_2$, $N(T_1) = (3; 0, 3, 0)$.

Thus the following composite DAs are obtained for part $A$: $C_1 = Q_1 \star T_1$, $C_2 = Q_2 \star T_1$.

Finally, the following eight resultant composite decisions are obtained:

$S_1 = A_1 \star B_1 \star C_1$, $S_2 = A_1 \star B_1 \star C_2$, $S_3 = A_1 \star B_2 \star C_1$, $S_4 = A_1 \star B_2 \star C_2$,

$S_5 = A_2 \star B_2 \star C_2$, $S_6 = A_2 \star B_1 \star C_2$, $S_7 = A_2 \star B_2 \star C_1$, and $S_8 = A_2 \star B_2 \star C_2$.

The resultant set of system solutions can be studied by the following ways: (1) multicriteria analysis, (2) expert judgment, and (3) additional usage of HMMD. Note in the example the initial combinatorial set includes 1179648 possible design solutions (i.e., $(2 \times 2 \times 4) \times (2 \times 3 \times 4) \times (4 \times 2) \times (3 \times 2) \times (2 \times 2 \times 2) \times (2 \times 2 \times 2)$).

### 3.3. Analysis and Improvement

Generally, improvement of composite DAs can be based on two kinds of actions (e.g., [20], [22]): (i) improvement of element, (ii) improvement of compatibility between elements. Table 10 contains improvement examples: bottlenecks (by elements, by compatibility) and improvement actions for composite DAs (system component $E$).

Table 10. Bottlenecks, improvement actions

| Composite DAs | Bottlenecks | | Actions $w/r$ |
|---|---|---|---|
| | DAs | IC | |
| $E_1 = J_2 \star K_1 \star L_1$ | $L_1$ | | $2 \Rightarrow 1$ |
| $E_1 = J_2 \star K_1 \star L_1$ | $J_2$ | | $3 \Rightarrow 1$ |
| $E_2 = J_1 \star K_2 \star L_3$ | $J_1$ | | $2 \Rightarrow 1$ |
| $E_2 = J_1 \star K_2 \star L_3$ | $K_1$ | | $2 \Rightarrow 1$ |
| $E_2 = J_1 \star K_2 \star L_3$ | | $(J_1, L_3)$ | $1 \Rightarrow 3$ |
| $E_2 = J_1 \star K_2 \star L_3$ | | $(J_1, K_2)$ | $1 \Rightarrow 3$ |
| $E_2 = J_1 \star K_2 \star L_3$ | | $(K_2, L_3)$ | $1 \Rightarrow 3$ |

## 4. CONCLUSION

The paper has described our hierarchical approach to modular composition of management systems for smart homes. Clearly, it is reasonable to consider other design problems (e.g., redesign/adaptation of management systems and/or its parts/components). In the future it may be prospective to consider the following research directions: *1.* study of redesign (improvement) problems; *2.* examination of designing a system trajectory (i.e., multistage system design); *3.* analysis of on-line adaption problems for smart home systems; *4.* usage of design models while taking into account uncertainty (e.g., stochastic models, fuzzy sets); *5.* usage of AI techniques in design procedures; *6.* examination of other components/modules of considered management system for smart homes; and *7.* usage of the described application and modular design approach in engineering/CS education.